\documentclass[12pt]{article}
\usepackage{latexsym}
\usepackage{graphicx}% Include figure files
\usepackage{dcolumn}% Align table columns on decimal point
\usepackage{amsfonts}
\usepackage{amsmath}

\def \m{\mbox}
\def \be{\begin{equation}}
\def \ee{\end{equation}}
\def \bea{\begin{eqnarray}}
\def \eea{\end{eqnarray}}
\def \ba{\begin{array}}
\def \ea{\end{array}}

\def\cos{\mbox{cos}}
\def\sinh{\mbox{sinh}}
\def\cosh{\mbox{cosh}}
\def\tanh{\mbox{tanh}}

\def \f{\frac}

\def \ro{\ddot{\rm{o}}}

\def \vp{\varphi}
\def \p{\partial}

\def \D{\cal{D}}

\begin{document}
\vspace*{-.6in}
\thispagestyle{empty}
\baselineskip = 18pt

\vspace{.5in}
\vspace{.5in}
{\LARGE
\begin{center}
Hawking Radiation of Black Rings from Anomalies
\end{center}}

\vspace{1in}

\begin{center}

Bin Chen\footnote{bchen01@pku.edu.cn}
\\
\emph{Department of Physics, Peking University, Beijing 100871,
P.R. China} \vspace{1.5cm}

Wei He\footnote{weihe@itp.ac.cn}
\\
\emph{Institute of Theoretical Physics,Chinese Academy of
Sciences,Beijing 100080 , P.R. China\\
and Graduate School of Chinese Academy of Sciences,Beijing 100049
, P.R. China}
\end{center}
\vspace{1in}

\begin{center}
\textbf{Abstract}
\end{center}
\begin{quotation}
\noindent We derive Hawking radiation of 5-dimensional black rings
from gauge and gravitational anomalies using the method proposed
by Robinson and Wilczek. We find as in the black hole case, the
problem could reduce to a (1+1) dimensional field theory and the
anomalies result in correct Hawking temperature for neutral,
dipole and charged black rings.
\end{quotation}

\newpage

\pagenumbering{arabic}

\section{Introduction}
Hawking radiation has been one of the central issues in black hole
quantum mechanics since its discovery\cite{HAW1976}. Recently
Robinson and Wilczek proposed a new derivation of Hawking
radiation of black hole via anomalies cancellation\cite{RW0502}.
They treated the simplest Schwarzschild black hole and found that
the cancellation of the gravitational anomalies result in correct
Hawking temperature. Subsequent works extend this method to
charged Reissner-Nordstr$\ro$m black hole in \cite{IUW0602} by
including gauge anomalies, and to rotating Kerr black hole in
\cite{ROT06}. The other related works treating various black holes
could be found in \cite{ALL}.

In the recent study of higher dimensional gravity, people find
it is  possible to have another class of black objects in $D\geq 5$
dimesion: black rings. Unlike black holes, their horizon topology is not
spherical. Especially in five dimension
explicit solutions have been found for  neutral, charged, and
supersymmetric black rings with horizon topology $S^1\times S^2$.
It's interesting to see if Robinson-Wilczek method applies to black
rings, and we find it indeed works. This seems natural because in
some limit black rings become Myers-Perry black holes, and their
Hawking radiation is given in \cite{IMU0612} by Robinson-Wilczek
method. We find that for five dimensional black rings the scalar
field theory reduces to a $(1+1)$ dimensional free field theory near
the horizon and the cancellation of gravitational and gauge
anomalies give us the correct Hawking temperature.

We review Robinson-Wilczek method briefly in the next section and
present our results in section three. In section four we discuss
the relation between Robinson-Wilczek method and the tunneling
picture\cite{PW9907} to get the Hawking temperature.

\section{Hawking radiation and anomalies}

In a black hole background we must be careful to define the
quantum state of a field theory. If we choose Boulware state as
the vacuum state, that is, define positive frequency using
Schwarzschild time, then at the horizon the energy-momentum tensor
of ground state is divergent due to ingoing modes. One way to
avoid this problem is to choose Unruh vacuum by defining positive
frequency using Kruskal coordinate, then these problematic ingoing
modes in Boulware vacuum are removed because they are excited
states now.

Robinson and Wilczek take another viewpoint\cite{RW0502,IUW0602}:
at classical level we just discard these ingoing modes near the
horizon because they can't affect region outside the horizon. Then
the quantum field theory near the horizon is chiral and suffers
from the gravitational anomalies, and the gauge anomalies if gauge
fields are presented. In order to cancel the anomalies to preserve
the symmetries, additional fluxes/currents should be included.
These fluxes/currens are quantum effects of the classically
irrelevant ingoing modes. The condition of the anomalies
cancellation and the regularity condition at the horizon determine
 Hawking flux of  charge and energy-momentum.

For several kinds of  black rings which will be studied in the
next section, the near horizon field theory reduces to a $(1+1)$
dimensional free field theory with a background metric in the
following form: \be \label{metric1}ds^2=-f(r)dt^2+\f{1}{f(r)}dr^2
\ee with $f(r_H)=0$ at the horizon $r_H$. The quantum field theory
in this small region $r_H\le r \le r_H+\epsilon$ could be treated
as chiral theory, and  suffers from anomalies. The gravitational
anomalies in two dimension take a simple form \be
\label{gravano}\nabla_\mu T_{\chi\nu}^\mu
=-\f{1}{96\pi\sqrt{-g}}\epsilon^{\beta\delta}\p_\delta
\p_\alpha\Gamma_{\nu\beta}^\alpha=\f{1}{\sqrt{-g}}\p_\mu N_\nu^\mu
=\cal B_\nu.\ee If there are background gauge fields, there exist
gauge anomalies  \be \label{gaugano}\nabla_\mu
J_\chi^\mu=\pm\f{e^2}{4\pi\sqrt{-g}}\epsilon^{\mu\nu}\p_\mu A_\nu.
\ee

As the symmetries of the underlying theory: general covariance and
gauge symmetry must be preserved, the anomalies in the region
$r_H\le r \le r_H+\epsilon$ must be cancelled by extra
currents/fluxes.

First we analyze the gauge anomalies. Introduce a step functions
$\Theta_+=\Theta(r-r_H-\epsilon )$ and $H=1-\Theta_{+}$ we can
write the current as\footnote{We don't take into account the
region inside the horizon because it's causally disconnected.} \be
J^i=J_{(H)}^iH+J_{(o)}^i\Theta_+. \ee Because we consider the
stationary configurations we have conserved current \be
\p_rJ_{(o)}^r=0 \ee and the anomalous current \be
\p_rJ_{(H)}^r=\f{e^2}{4\pi}\p_rA_t. \ee They are solved by\be
J_{(o)}^y=c_o \qquad J_{(H)}^y=c_H+\f{e^2}{4\pi}(A_t(r)-A_t(r_H))
\ee where $c_o,c_H$ are integration constants. The $c_o$ is the
value of the flux at infinity and $c_H$ is the value of the
current of the outgoing modes at the horizon. The gauge invariance
of the full theory demands \be -\delta_\lambda W=\int
d^2x\lambda[(J_{(o)}^r-J_{(H)}^r+\f{e^2}{4\pi}A_t)
\delta(r-r_H-\epsilon)+\p_r(\f{e^2}{4\pi}A_tH)]=0\ee where
$\lambda$ is the gauge parameter. The last term in the integral is
cancelled by the quantum effects of the classically irrelevant
ingoing modes. In the limit $\epsilon\to0$ we have
 \be
c_o=c_H-\f{e^2}{4\pi}A_t(r_H) \ee In order to fix the constants we
impose a boundary condition: the covariant current
$\tilde{J}^i=J^i+\f{e^2}{4\pi}A_t(r)H$ vanish at the horizon. This
is the regularity condition as pointed out in the first reference
of \cite{ROT06}. Then the integration constants are fixed\be
c_o=2c_H=-\f{e^2}{2\pi}A_t(r_H).\ee It represents  Hawking flux of
charge.

For the gravitational anomalies, we can write $T_\nu^\mu$ as \be
T_\nu^\mu=T_{(H)\nu}^\mu H+T_{(o)\nu}^\mu \Theta_{+}\ee
For the metric (\ref{metric1}) we have : \bea \nonumber N_t^t&=&N_r^r=0\\
\nonumber
N_t^r&=&\f{1}{192\pi}(ff^{''}+f^{'2})\\
N_r^t&=&\f{1}{192\pi}\f{f^{''}f-f^{'2}}{f^2} \eea where prime
denotes derivative with respect to $r$.

Under diffeomorphism transformation $x\to x^{'}=x-\xi$, taking
gauge fields into account we have \be -\delta_\xi W=\int
d^2x\sqrt{-g_{(2)}}\xi^\nu\nabla_\mu T_\nu^\mu \ee where \be
\nabla_\mu T_{\nu}^{\mu}=F_{\mu\nu}J^\mu+A_\nu\nabla_\mu
J^\mu+{\cal{B}}_\nu \ee For the energy-momentum tensor in radial
direction, in the region $r>r_H+\epsilon$ we have
 \be \p_rT_{(o)t}^r=F_{rt}J_{(o)}^r;\ee
in the region $r_H \le r\le r_H+\epsilon$, since $\nabla_\mu
J^\mu\ne0,N_t^r\ne0$, we have \be
\p_rT_{(H)t}^r=F_{rt}J_{(H)}^r+A_t\p_rJ_{(H)}^r+\p_rN_t^r. \ee
They are solved by \bea\nonumber
T_{(o)t}^r&=&a_o+c_oA_t(r)\\
T_{(H)t}^r&=&a_H+\int_{r_H}^r\p_r(c_oA_t+\f{e^2}{4\pi}A_t^2+N_t^r)dr
\eea The general covariance demands \be \int
d^2x\xi^t[\p_r(A_tJ_{(H)}^rH)+\p_r(N_t^rH)
+(T_{(o)t}^r-T_{(H)t}^r-A_tJ_{(H)}^r+N_t^r)\delta(r-r_H-\epsilon)]=0
\ee The first term in the integral should be cancelled by the
quantum effect of the ingoing modes. Taking  $\epsilon\to 0$
limit, we get \be a_o=a_H+\f{e^2}{4\pi}A_t^2(r_H)-N_t^r(r_H) \ee
Imposing a vanishing condition for the covariant energy-momentum
$\tilde{T}_t^r=T_t^r+(ff^{''}-2f^{'2})/192\pi$ on the horizon, we
get \be a_H=2N_t^r(r_H)\qquad
a_o=\f{e^2}{4\pi}A_t^2(r_H)+N_t^r(r_H). \ee The $a_o$ represents
Hawking flux of energy-momentum.

 The flux
$\Phi=N_t^r(r_H)$ represents a thermal flux due to Hawking
radiation, it's related to temperature as $\Phi=\f{\pi}{12}T^2$.
Therefore  we can get Hawking temperature \be
T=\f{f^{'}(r_H)}{4\pi} \ee

\section{Hawking radiation of 5-dimensional black rings from anomalies}
\subsection{Neutral black ring}
The five dimensional neutral black ring was first found in
\cite{ER0110} as a vacuum solution of five dimensional general
relativity. Black rings obey similar thermodynamics laws as black
holes, while lose uniqueness because in some range of their
parameter space there are more than one solution with the same
conserved charges. For more aspects about five dimensional black
rings, see the review \cite{ERreview}.

The metric of the neutral black ring is (we use the coordinates
given in \cite{E0402})
\bea \nonumber ds^2&=&-\f{F(y)}{F(x)}(dt-CR\f{1+y}{F(y)}d\psi)^2\\
 &\quad&+\f{R^2}{(x-y)^2}F(x)[-\f{G(y)}{F(y)}d\psi^2-\f{dy^2}{G(y)}+\f{dx^2}{G(x)}
+\f{G(x)}{F(x)}d\phi^2] \eea with functions \be F(\xi)=1+\lambda\xi
\qquad G(\xi)=(1-\xi^2)(1+\nu\xi) \ee and constant \be
C=\sqrt{\lambda(\lambda-\nu)\f{1+\lambda}{1-\lambda}},\qquad
0<\nu\le\lambda<1 \ee the coordinates $\psi,\phi$ are two cycles of
the black ring and $x,y$ take values \be -1\le x\le 1,\qquad
-\infty\le y\le -1 \ee

The center of the black ring is located at $y=-\infty$, and $x\to
-1,y\to-1$ approaches the asymptotic infinity. The horizon is at
$y=y_H=-\f{1}{\nu}$ with the topology $S^1\times S^2$. The mass of
the black ring is \be M=\f{3\pi R^2}{4G}\f{\lambda}{1-\nu} \ee In
order to prevent the contraction the ring must rotate along the
$\psi$ direction, the angular momentum is\be J=\f{\pi
R^3}{2G}\f{\sqrt{\lambda(\lambda-\nu)(1+\lambda)}}{(1-\nu)^2} \ee
The neutral black rings are parameterized by their mass and
angular momentum, or by $\lambda$ and $\nu$.

Now let us show that near the horizon a scalar field theory
reduces to a (1+1) dimensional free field theory. For black ring
metric we have \be \sqrt{-g}=\f{R^4F(x)}{(x-y)^4} \ee and the
non-zero components of the inverse metric are \bea \nonumber
g^{tt}&=&-\f{C^2(x-y)^2(1+y)^2}{F(x)F(y)G(y)}-\f{F(x)}{F(y)}\qquad
g^{t\psi}=-\f{C(x-y)^2(1+y)}{RF(x)G(y)}\\ \nonumber
g^{\psi\psi}&=&-\f{(x-y)^2F(y)}{R^2F(x)G(y)}\qquad
g^{xx}=\f{(x-y)^2G(x)}{R^2F(x)}\\
g^{yy}&=&-\f{(x-y)^2G(y)}{R^2F(x)}\qquad
g^{\phi\phi}=\f{(x-y)^2}{R^2G(x)} \eea For a scalar field $\vp$ in
the black ring background, the action is \bea \nonumber
S[\vp]&=&\int
d^5x(\vp\p_i\sqrt{-g}g^{ij}\p_j\vp+\sqrt{-g}V_{int}(\vp))\\
\nonumber&=&\int
d^5x\vp[-(\f{C^2R^4(1+y)^2}{(x-y)^2F(y)G(y)}+\f{R^4F(x)^2}{(x-y)^4F(y)})\p_t^2
-2\f{CR^3(1+y)}{(x-y)^2G(y)}\p_t\p_\psi\\
\nonumber&\qquad&-\f{R^2F(y)}{(x-y)^2G(y)}\p_\psi^2
+\p_x\f{R^2G(x)}{(x-y)^2}\p_x
-\p_y\f{R^2G(y)}{(x-y)^2}\p_y+\f{R^2F(x)}{(x-y)^2G(x)}\p_\phi^2]\vp\\
&\qquad&+\int d^5x \f{R^4F(x)}{(x-y)^4}V_{int}(\vp) \eea We expand
the scalar field as
\be\label{expan}\vp=\sum_{k,l}\f{1}{2\pi}\vp^{(kl)}(t,x,y)e^{ik\phi}e^{il\psi}\ee
where $k,l$ are integers because $\phi,\psi$ are periodic
coordinates. Put it into the action and take the near horizon
limit $y\to y_H,G(y)\to 0$, leaving the dominant terms in the
action we get \bea  S[\vp]&=&\int
d^5x\vp^{(kl)}\f{R^2}{(x-y)^2}[-\f{C^2R^2(1+y)^2}{F(y)G(y)}(\p_t+i\f{l
F(y)}{CR(1+y)})^2-\p_yG(y)\p_y]\vp^{(kl)} \label{rea}.\eea

As in black hole cases, the potential term for the scalar field is
suppressed. Moreover, the terms involving $\p_x$ are also
suppressed and do not appear in the action. Therefore we can
further expand function $\vp^{(kl)}(t,x,y)$ in terms of $x$. As
$x=\cos\theta$ in polar coordinate, an appropriate expansion
function is the Legendre polynomial $P_n(x)$,  \be
\vp^{(kl)}(t,x,y)=\sum_n \vp^{(kln)}(t,y)P_n(x). \ee  Moving
$P_n(x)$ to the left of the operator, we integrate over $x$ and
have : \be \int_{-1}^{1}dx\f{P_m(x)P_n(x)}{(x-y)^2}=a_{mn}(y) \ee
Then the action can be written as \bea \nonumber S[\vp]&=&\int
dtdy\sum_{k,l,m,n}a_{mn}(y)\f{CR^3(1+y)}{\sqrt{-F(y)}} \times
\vp^{(klm)}(t,y)\\&\quad&[\f{CR(1+y)}{\sqrt{-F(y)}G(y)}(\p_t+il\f{
F(y)}{CR(1+y)})^2
-\p_y\f{\sqrt{-F(y)}}{CR(1+y)}G(y)\p_y]\vp^{(kln)}(t,y) \eea

In the above action we can treat $\f{F(y)}{CR(1+y)}$ as an
effective gauge field $A_t(y)$ and $l$ serves as the gauge
coupling constant. The coefficient $a_{mn}(y)$ is symmetric about
$m,n$. We can further absorb the factor
$a_{mn}(y)\f{CR^3(1+y)}{\sqrt{-F(y)}}$ in to $\vp^{(kln)}(t,y)$
and define a new field $\tilde{\vp}^{(kln)}(t,y)$, then the action
can be written in a canonical form \bea \label{2daction} \nonumber
S[\vp]&=&\int
dtdy\sum_{k,l,n}\tilde{\vp}^{(kln)}(t,y)\\&\quad&\times
[\f{CR(1+y)}{\sqrt{-F(y)}G(y)}(\p_t+il\f{F(y)}{CR(1+y)})^2
-\p_y\f{\sqrt{-F(y)}}{CR(1+y)}G(y)\p_y]\tilde{\vp}^{(kln)}(t,y)
\eea It's clear we get an infinite set of effective free massless
fields $\tilde{\vp}^{(klm)}(t,y)$ in (1+1) dimension with metric
\be \label{metric2}ds^2=-f(y)dt{^2}+\f{1}{f(y)}dy^2 \ee where \be
f(y)=\f{\sqrt{-F(y)}}{CR(1+y)}G(y) \ee together with a background
$U(1)$ gauge field\be A_t(y)=-\f{F(y)}{CR(1+y)} \ee In the near
horizon region $y=-\f{1}{\nu}+\epsilon$ we have
$F(y)<0,G(y)<0,(1+y)<0$, so $t$ is timelike and $y$ is spacelike.

It is remarkable that the essential point in the above discussion
is the absence of the derivatives with respect to $x$ in the
reduced action (\ref{rea}). This indicates that there is no
dynamics in $x$ direction. Effectively we may take $x$ as just a
parameter to label the field. This is the reason why we get a
two-dimensional free scalar field theory in the near horizon
limit.

The analysis of the gauge and gravitational anomalies in this two
dimensional theory is the same as in section 2. For the gauge
anomalies we have \be J_{(o)}^y=c_o \qquad
J_{(H)}^y=c_H+\f{l^2}{4\pi}(A_t(y)-A_t(y_H)) \ee with \be
c_o=2c_H=-\f{l^2}{2\pi}A_t(y_H)=\f{l^2}{2\pi}\Omega_H. \ee We have
written $-A_t(y_H)$ as $\Omega_H$ where $\Omega_H$ is the angular
velocity at the horizon.

For the gravitational anomalies we have \bea\nonumber
T_{(o)t}^y&=&a_o+c_oA_t(y)\\
T_{(H)t}^y&=&a_H+(c_oA_t+\f{l^2}{4\pi}A_t^2+N_t^y)|_{y_H}^y \eea
with \be a_o=\f{l^2}{4\pi }\Omega_H^2+N_t^y(y_H)\qquad
a_H=2N_t^y(y_H) \ee

From the relation $\Phi=\f{\pi}{12}T^2$, where $\Phi$ is the flux
$N_t^y(y_H)$, we get the Hawking temperature of the black ring \be
T=\f{f^{'}(y_H)}{4\pi}=\f{1}{4\pi
R}\f{1+\nu}{\sqrt{\lambda\nu}}\sqrt{\f{1-\lambda}{1+\lambda}}. \ee

\subsection{Dipole black ring}

The five dimensional dipole black rings was first constructed in
\cite{E0402}, its metric is of the form\bea \nonumber
ds^2&=&-\f{F(y)H(x)}{F(x)H(y)}(dt-CR\f{1+y}{F(y)}d\psi)^2\\
&+&\f{R^2F(x)H(x)H(y)^2}{(x-y)^2}[-\f{G(y)}{F(y)H(y)^3}d\psi^2-\f{dy^2}{G(y)}+\f{dx^2}{G(x)}
+\f{G(x)}{F(x)H(x)^3}d\phi^2] \eea with $F(\xi),G(\xi)$ being
defined as before, and the new function $H(\xi)$ being: \be
H(\xi)=[H_1(\xi)H_2(\xi)H_3(\xi)]^{\f{1}{3}}\quad\m{with}\quad
H_i(\xi)=1-\mu_i\xi \ee where $0\le\mu_i<1, i=1,2,3$. The $\mu_i$'s
are parameters related to the dipoles of  black ring. The scalar and
gauge fields are \be X^i=\f{H(x)H_i(y)}{H(y)H_i(x)} \qquad
A_i=C_iR\f{1+x}{H_i(x)}d\phi \ee where $C_i$'s are of the same form
as $C$ but with $\lambda\to -\mu_i$. As the gauge fields are
magnetic, they represent circularly distributed monopole charges.
This kind of black ring has no conserved gauge charges while allows
continuous value of dipoles: \be q_i=\f{1}{2\pi}\int_{S^2}dA^i \ee
so dipole black rings continuously violate the uniqueness. These
dipole charges are not conserved charges of black rings, but they do
appear in the first law of thermodynamics of black
rings\cite{E0402,TERMO}.

We can analyze the scalar action in this dipole black ring
background and expand $\vp$ in the same way as in the neutral
case, after field redefinition we can get a two dimensional
massless free scalar field theory with action similar with
(\ref{2daction}). We will not list the details but just give the
final form: \bea \nonumber S[\vp]&=&\int dtdy\sum_{k,l,n}
\tilde{\vp}^{(kln)}(t,y)[\f{CR(1+y)H(y)^{3/2}}{\sqrt{-F(y)}G(y)}\\&\quad&(\p_t+\f{
ilF(y)}{CR(1+y)})^2
-\p_y\f{\sqrt{-F(y)}G(y)}{CR(1+y)H(y)^{3/2}}\p_y]\tilde{\vp}^{(kln)}(t,y).
\eea Note that gauge fields $A_{i\phi}$ do not contribute dominant
terms in the action. The action is similar to the neutral case
except a new $H(y)^{3/2}$ factor.

The back ground metric is of the same form as (\ref{metric2}) with
\be f(y)=\f{\sqrt{-F(y)}}{CR(1+y)H(y)^{3/2}}G(y) \ee and together
with a gauge field \be A_t(y)=-\f{ F(y)}{CR(1+y)} \ee

Note that $A_{i\phi}$'s are not contained in our (1+1) dimensional
theory because they are magnetic, and their electric dual are
two-form fields that do not couple to point particles. In fact,
$A_{i\phi}$'s behave as three scalar fields in the near horizon
2-dimensional background.

The analysis of both the gauge and the gravitational anomalies is
parallel with the neutral case except their $f(y)$'s are
different. The Hawking temperature of the dipole black ring is\be
T=\f{f^{'}(y_H)}{4\pi}=\f{1}{4\pi
R}\f{1+\nu}{\sqrt{\lambda\nu}}\sqrt{\f{1-\lambda}{1+\lambda}}
\f{1}{\sqrt{\prod_i(1+\f{\mu_i}{\nu})}} \ee

\subsection{Charged black ring}

The rotating black ring with a single electric charge was first
constructed in \cite{E0305} as a solution of low energy effective
action of heterotic string, black rings with two and three charges
were given in \cite{EE0310}. We only treat the single-charged
black ring in this subsection. For black rings with two or three
charges,  the discussion are similar because charged black rings
have very similar structure.

The metric of the single-charged ring can be written as
\footnote{Metric in this form can be obtained from D1-D5 solution
in 6-dimension given in apendix.A of \cite{EE0310} by a
dimensional reduction in $z$-direction and set charges equal
$\alpha_1=\alpha_5=\alpha$. In order to write it in the form
consistent with neutral and dipole cases a further coordinates
transformation is needed\cite{ERreview}:$ x\to
\f{x+\lambda}{1+\lambda x},y\to \f{y+\lambda}{1+\lambda
y},(\psi,\phi)=\f{1-\lambda\nu}{\sqrt{1-\lambda^2}}(\psi,\phi),\lambda\to
\lambda,\nu\to \f{\lambda-\nu}{1-\lambda\nu}. $ }
\bea \nonumber ds^2&=&-\f{F(y)}{F(x)K(x,y)^2}(dt-CR\f{1+y}{F(y)}\cosh^2\alpha d\psi)^2\\
 &\quad&+\f{R^2}{(x-y)^2}F(x)[-\f{G(y)}{F(y)}d\psi^2-\f{dy^2}{G(y)}+\f{dx^2}{G(x)}
+\f{G(x)}{F(x)}d\phi^2] \eea where functions $F(\xi),G(\xi)$ are
defined as before, and $K(x,y)$ is defined as \be
K(x,y)=1+\f{\lambda(x-y)}{F(x)}\sinh^2\alpha \ee where $\alpha$ is
the parameter represents the electric charge.

The dilation field is \be e^{-\Phi}=K(x,y) \ee and the gauge
fields are \be
A_t=\f{\lambda(x-y)\sinh\alpha\cosh\alpha}{F(x)K(x,y)}, \qquad
A_\psi=\f{CR(1+y)\sinh\alpha\cosh\alpha}{F(x)K(x,y)} \ee with
electric charge \be
Q=\f{2\sinh2\alpha}{3(1+\f{4}{3}\sinh^2\alpha)}M \ee there is also
a 2-form $B_{t\psi}$ field which indicates that the black ring
carries local fundamental string charge, but it is irrelevant to
our discussion because it does not couple to point particles.

The action of a scalar field in this black ring background is \be
S[\vp]=\int
d^5x(\vp{\D}_i\sqrt{-g}g^{ij}{\D}_j\vp+\sqrt{-g}V_{int}(\vp)). \ee
The terms containing covariant derivatives appear as \be
\f{R^2}{(x-y)^2K(x,y)}[-\f{C^2R^2(1+y)^2\cosh^4\alpha}{F(y)G(y)}{\D}_t^2
-2\f{CR(1+y)\cosh^2\alpha}{G(y)}{\D}_t{\D}_\psi-\f{F(y)}{G(y)}{\D}_\psi^2].
\ee We expand $\vp$ as (\ref{expan}) and put it into the action,
after taking near horizon limit $y\sim y_H+\epsilon$ we get the
dominant action for field $\vp^{(kl)}(t,x,y)$: \bea \nonumber
S[\vp]&=&\int
d^5x\vp^{(kl)}\f{R^2}{(x-y)^2K(x,y)}[-\f{C^2R^2(1+y)^2\cosh^4\alpha}{F(y)G(y)}
(\p_t+ieA_t\\\nonumber&\quad&+\f{ieA_\psi
F(y)}{CR(1+y)\cosh^2\alpha}+\f{ilF(y)}{CR(1+y)\cosh^2\alpha})^2
-\p_yG(y)\p_y+\p_xG(x)\p_x]\vp^{(kl)}\\ \nonumber &=&\int
d^5x\vp^{(kl)}\f{R^2}{(x-y)^2K(x,y)}[-\f{C^2R^2(1+y)^2\cosh^4\alpha}{F(y)G(y)}
\\&\quad&(\p_t+ie\tanh\alpha+\f{ilF(y)}{CR(1+y)\cosh^2\alpha})^2
-\p_yG(y)\p_y]\vp^{(kl)}. \eea In the second step we discard
$\p_x$ part because the $x$-dependence coming from $A_t(x,y)$ and
$A_\psi(x,y)$ cancels exactly.

Now as the operator in the bracket does not depend on $x$ anymore,
we can expand $\vp^{(kl)}(t,x,y)$ as before
$\vp^{(kl)}(t,x,y)=\sum_n\vp^{(kln)}(t,y)P_n(x)$. After
integrating $x$ \be \int_{-1}^1
dx\f{P_m(x)P_n(x)}{(x-y)^2K(x,y)}=b_{mn}(y), \ee  the action
becomes \bea \nonumber S[\vp]&=&\int
dtdy\f{CR^3(1-y)\cosh^2\alpha} {\sqrt{-F(y)}}b_{mn}(y)
\vp^{(klm)}(t,y)
[\f{CR(1+y)\cosh^2\alpha}{\sqrt{-F(y)}G(y)}\\
&&(\p_t+ie\tanh\alpha+il\f{F(y)}{CR(1+y)\cosh^2\alpha})^2
-\p_y\f{\sqrt{-F(y)}G(y)}{CR(1+y)\cosh^2\alpha}\p_y]
\vp^{(kln)}(t,y). \eea After field redefinition,  we can write the
action in the canonical form. The action describe an infinite
collection of massless scalar fields in a (1+1) dimensional
background, whose metric is of the form (\ref{metric2}) with \be
f(y)=\f{\sqrt{-F(y)}}{CR(1+y)\cosh^2\alpha}G(y),\ee  and coupling
to two $U(1)$ fields \be A_{t}^{(1)}=-\tanh\alpha,\qquad
A_{t}^{(2)}(y)=-\f{F(y)}{CR(1+y)\cosh^2\alpha}.\ee Note that
$A_t^{(2)}(y_H)=-\Omega_H$ where $\Omega_H$ is the angular
velocity at the horizon.

Extend the gauge anomalies analysis to more than one gauge fields is
straightforward. The anomaly equations for electric currents
are\bea\nonumber
\p_yJ_{(H)}^{(1)y}&=&\f{e}{4\pi}\p_y(eA_t^{(1)}+lA_t^{(2)})\\
\p_yJ_{(H)}^{(2)y}&=&\f{l}{4\pi}\p_y(eA_t^{(1)}+lA_t^{(2)}) \eea
As $A_{t}^{(1)}$ is a constant it in fact does not contribute
anomalies in the region $y_H\le y\le y_H+\epsilon$, so we have \be
J_{(o)}^{(1)y}=c_{o}^{(1)} \qquad
J_{(H)}^{(1)y}=c_{H}^{(1)}+\f{el}{4\pi}(A_t^{(2)}(y)-A_t^{(2)}(y_H)).\ee
The gauge invariance of $A_t^{(y)}$ demands \be
c_{o}^{(1)}=c_{H}^{(1)}-\f{e}{4\pi}(eA_t^{(1)}(y_H)+lA_t^{(2)}(y_H))
\ee and the regular boundary condition leads to\be\label{fluxe}
c_{o}^{(1)}=\f{e}{2\pi}(e\tanh\alpha+l\Omega_H) \ee Similarly we
can get $J^{(2)y}$.

For gravitational anomalies we have \bea\nonumber
T_{(o)t}^y&=&a_o+c_{o}^{(1)}A_{t}^{(1)}(y)+c_{o}^{(2)}A_{t}^{(2)}(y)\\
T_{(H)t}^y&=&a_H+\int_{y_H}^y\p_y(c_{o}^{(2)}A_t^{(2)}+\f{l^2}{4\pi}A_{t}^{(2)2}+N_t^y)dy.
\eea The general covariance demands \be
a_o=a_H-c_{H}^{(1)}A_{t}^{(1)}(y_H)-c_{H}^{(2)}A_{t}^{(2)}(y_H)-N_t^y(y_H)\ee
A covariance boundary condition fixes \be
\label{fluxT}a_H=2N_t^y(y_H),\qquad
a_o=\f{1}{4\pi}(e\tanh\alpha+l\Omega_H)^2+N_t^y(y_H) \ee

The Hawking temperature is given by \be
T=\f{f^{'}({y_H})}{4\pi}=\f{1}{4\pi
R\cosh^2\alpha}\f{1+\nu}{\sqrt{\lambda\nu}}\sqrt{\f{1-\lambda}{1+\lambda}}.
\ee In the extremal limit $\alpha\to\infty,$, we have $Q\to M,T\to
0$.

Now let's compare our results with the black body radiation of
black rings at temperature $T=\f{1}{\beta}$ , in order to avoid
superradiance problem we only consider fermions. The Plank
distribution of fermions with energy $\omega$, charge $e$ and
angular momentum $l$ is \be
N_{e,l}(\omega)=\f{1}{e^{\beta(\omega-e\Phi-l\Omega_H)}+1} \ee
where $\Phi$ is the co-rotating electric chemical potential at the
horizon and $\Omega_H$ is the angular velocity at the horizon. The
Killing vector of black ring background is
$\xi=\p_t+\Omega_H\p_\psi$, so we have\be \Phi=\xi^\mu
A_\mu=A_t(x,y_H)+\Omega_HA_\psi(x,y_H)=\tanh\alpha. \ee Then the
Hawking flux of electric charge, angular momentum and
energy-momentum tensor are\bea \nonumber J_e&=&e\int_0^\infty
\f{d\omega}{2\pi}(N_{e,l}(\omega)-N_{-e,-l}(\omega))=\f{e}{2\pi}(e\tanh\alpha+l\Omega_H)\\\nonumber
J_l&=&l\int_0^\infty
\f{d\omega}{2\pi}(N_{e,l}(\omega)-N_{-e,-l}(\omega))=\f{l}{2\pi}(e\tanh\alpha+l\Omega_H)\\
J_E&=&\int_0^\infty
\f{d\omega}{2\pi}\omega(N_{e,l}(\omega)+N_{-e,-l}(\omega))=\f{1}{4\pi}(e\tanh\alpha+l\Omega_H)^2+\f{\pi}{12\beta^2}.
\eea They agree with the results (\ref{fluxe},\ref{fluxT}) derived
from anomalies cancellation. Set $e=0$ we get the results for
neutral black ring.

\section{Relation to tunneling picture}
In \cite{PW9907} Parrikh and Wilczek showed how to derive Hawking
temperature in the tunneling picture. We need to calculate the
tunneling probability of a particle tunnels from
$r_{in}=r_H(M)-\epsilon$ inside the initial  horizon to
$r_{out}=r_H(M-\omega)+\epsilon$ outside the final horizon in the
Painleve coordinate. Under WKB approximation, it is:\be \Gamma\sim
e^{-2\m{Im}S} \ee with \be
\m{Im}S=\m{Im}\int_{r_{in}}^{r_{out}}p_rdr
=\m{Im}\int_0^{\omega}\int_{r_{in}}^{r_{out}}\f{dr}{\dot{r}}d(-\omega).\ee
Here $\omega$ is the energy of the radiated particle,
$M(M-\omega)$ is the mass of  black hole before (after) emission.
And $\dot{r}$ is the radial null geodesic, it's a function of $r$
and conserved charges such as $M,Q,J$. As $\omega\ll M$, to first
order the above expression is approximately:\be
\label{tunneling}\m{Im}S\simeq \omega
\m{Im}\int_{r_{out}}^{r_{in}}\f{dr}{\dot{r}} \ee

Let us apply the above analysis to the black rings studied above.
For simplicity, we focus on the neutral black ring and consider
the effective two-dimensional near-horizon metric (\ref{metric2})
obtained in Robinson-Wilczek method.
%Let's see the two dimensional
%metric (\ref{metric2}) coming from neutral black ring.
Transforming $t$ coordinate as \be t\to t-\int
\f{CR(1+y)}{\sqrt{-F(y)}G(y)}\sqrt{1-G(y)}dy \ee we can rewrite
the metric in the Painleve form
 \be
ds^2=-\f{\sqrt{-F(y)}G(y)}{CR(1+y)}dt^2+2\sqrt{1-G(y)}dtdy
+\f{CR(1+y)}{\sqrt{-F(y)}}dy^2
 \ee
Consider a uncharged particle to avoid the influence of
electromagnetic fields, the null geodesics is \be
\dot{y}=\f{\sqrt{-F(y)}}{CR(1+y)}(\pm 1-\sqrt{1-G(y)}) \ee where
$\pm$ correspond to ougoing/ingoing geodesics.

Performing the integral (\ref{tunneling}) with outgoing geodesic, we
have \bea \nonumber \m{Im}S&\simeq&
\omega\f{CR(1-\f{1}{\nu})}{\sqrt{-F(-\f{1}{\nu})}}\times
2\pi\m{Res}_{y_H}\f{1}{1-\sqrt{1-G(y)}}\\
&\simeq& \omega\f{CR(1-\f{1}{\nu})}{\sqrt{-F(-\f{1}{\nu})}}\times
2\pi\f{2\nu}{\nu^2-1}.\eea From $\Gamma\sim e^{-2\m{ImS}}\sim
e^{-\f{\omega}{T}}$, we get the Hawking temperature \be
T=\f{1}{4\pi
R}\f{1+\nu}{\sqrt{\lambda\nu}}\sqrt{\f{1-\lambda}{1+\lambda}}. \ee
The situations for dipole and charged black rings are similar.

\section{Conclusion} In this paper, we extended Robinson-Wilczek method to the five
dimensional black rings and got their Hawking temperature and
fluxes correctly. As in the rotating black hole case, near the
horizon the field theory reduces to a (1+1) dimensional free field
theory coupled to gauge fields after appropriate field
redefinition. From the cancellation of the gauge and gravitational
anomalies, we obtain the correct Hawking temperature and
radiation. The same two dimensional metric also determines the
null geodesics in the tunneling method and leads to the same
Hawking temperature. This reflects the fact that the Hawking
radiation is
 determined universally by the horizon properties.

From studying various black holes and black rings here, we expect
for stationary black objects in various dimensions with more
complicated horizon topology the Robinson-Wilczek method still
works. The horizon is a hypersurface with uniform physical
properties, for example the surface gravity and electric potential
are constant on the horizon. Our study of black rings shows that
quantum field theory near the horizon is relatively simple. In the
near horizon limit, $G(y)$ captures the singular structure. After
zooming in the horizon, the horizon is effectively a
two-dimensional one and the quantum field theory is effectively
coupled to the gravitational background (\ref{metric2}) after
field redefinition. From the analysis of charged black ring we see
that near the the horizon the electrodynamics also simplifies,
depending only on the radial direction, while outside the horizon
it's much more complicated.

The Robinson-Wilczek method has been tested in many cases, at
present it does not deal with other aspects of black hole physics
such as thermodynamics and entropy, a better understanding on
these issues is expected. At least for studying Hawking radiation
with Robinson-Wilczek method, we should expect for more general
cases the near horizon physics could be simplified and reduced to
$(1+1)$ dimensional one, if the horizon is only controlled by a
function depending on one coordinate. Though the exploration is
still case by case, it seems that this
 method is quite general.

\section*{Acknowledgments}

The work was partially supported by NSFC Grant No.
10405028,10535060, NKBRPC (No. 2006CB805905) and the Key Grant
Project of Chinese Ministry of Education (NO. 305001).\\
\\{\sl Note added:}After submit this note we are informed an
independent work by Umpei Miyamoto and Keiju Murata \cite{MM0705}
which has some overlap with our work.

\end{document}